# ELECTROMAGNETIC FIELD PARTICLES IN THE CLASSICAL THEORY

S.Botrić[*] and K.Ljolje
Academy of Sciences and Arts of Bosnia and Herzegovina,
Sarajevo, Bosnia and Herzegovina

***Abstract***

Appearance of particles like photons within the classical electromagnetic field theory is demonstrated.

---

[*] Faculty of Electrical Engineering, Mechanical Engineering and Naval Architecture, University of Split, Split, Croatia

## 1. Introduction

Einstein s hypothesis of photons has followed the Planck's quantum hypothesis. This hypothesis tells that the electromagnetic field is composed of particles-photons-with the energy-momentum fourvector

$$P^\alpha = \mathrm{h}k^\alpha. \qquad (1.1)$$

Light quantum hypothesis concerns the energy of a monochromatic plane wave as well as the existence of a relativistic scalar h.

M. Abraham [1] was first to derive for the special case of circularly polarized monochromatic plane wave the ratio between the spin angular momentum and the time average of the total energy of the plane wave in the classical description of electromagnetic waves.

This ratio indicates the existence of a scalar constant having the dimension of angular momentum. One could infer that this ratio would be in full accordance with light quantum hypothesis if the scalar constant equals $1/2\pi$ times $h$.

Concerning invariance under proper homogeneous Lorentz transformations of the free electromagnetic field with the Lagrangian

$$L_{elm} = -\frac{1}{16\pi} F_{\mu\nu} F^{\mu\nu} \qquad (1.2)$$

F. Rohrlich [2] finds in accord with [1] that in certain special cases of plane waves the aforesaid ratio is given by

$$\frac{\Sigma}{\overline{W}} = \frac{\underline{k}}{k} \frac{2ab}{a^2 + b^2} \frac{\sin \delta}{\omega}, \tag{1.3}$$

where a, b, δ, $\underline{k}$, ω are parameters of the plane wave. In connection with this formula Rohrlich says: "The classical result (4-188) can be combined with Max Planck s light quantum hypothesis of the year 1900. If $\overline{W} = \eta\omega$ where η is 1/2π times Planck's constant h, then right and left circularly polarized light must have a spin angular momentum of $\Sigma = h\underline{k}$ and $-h\underline{k}$, respectively. The quantum field theoretical description of electromagnetic waves confirms this conclusion".

In [1] and [2] no scalar constant of the free electromagnetic field has been determined and consequently these considerations gave a partial insight into photon from the view-point of classical electromagnetic field theory.

In the article [3] the scalar constant of the free electromagnetic field and its consequences have been analyzed. In this article we show that the whole concept of the photon is potentially present in the conceptual framework of the classical electromagnetic field theory.

## 2. *A Decomposition of Electromagnetic Plane Wave*

In the case of the free electromagnetic field with the Lagrangian density

$$L_{em} = -\frac{1}{16\pi} F_{\mu\nu} F^{\mu\nu}, \tag{2.1}$$

where are

$$F_{\mu\nu} = \partial_\mu A_\nu - \partial_\nu A_\mu, \tag{2.2}$$

$$A^\mu = \left(A^0, \underline{A}\right),\tag{2.3}$$

a plane wave solution of the equation (2.4)

$$\partial_\mu F^{\mu\nu} = 0 \tag{2.4}$$

can be decomposed into the two plane wave solutions:

$$F^{\mu\nu} = F_1^{\mu\nu} + F_2^{\mu\nu},\tag{2.5}$$

$$F_1^{\mu\nu} = \partial^\mu A_1^\nu - \partial^\nu A_1^\mu,\tag{2.6}$$

$$F_2^{\mu\nu} = \partial^\mu A_2^\nu - \partial^\nu A_2^\mu,\tag{2.7}$$

$$\partial_\mu F_i^{\mu\nu} = 0 \quad (i = 1, 2).\tag{2.8}$$

Since the three-vectors $\underline{E}, \underline{B}$ of electromagnetic plane wave propagating in the direction defined by the unit vector $\underline{n}$ satisfy the relation

$$\underline{B} = \underline{n} \times \underline{E},\tag{2.9}$$

the decomposition (2.5-2.8) can be in Coulomb gauge defined more specifically in terms of the three-vectors $\underline{A}_1$ and $\underline{A}_2$, where $A^\mu = (0, \underline{A})$, $A_1^\mu = (0, \underline{A}_1)$, $A_2^\mu = (0, \underline{A}_2)$ are the four-vectors of electromagnetic potentials:

$$\underline{A} = \underline{A}_1 + \underline{A}_2,\tag{2.10}$$

$$\underline{A}_1 \times \underline{A}_2 = a^2 \underline{n}, \tag{2.11}$$

$$\underline{A}_2 \times \underline{n} = \underline{A}_1, \tag{2.12}$$

$$\underline{n} \times \underline{A}_1 = \underline{A}_2, \tag{2.13}$$

and $a$ is a real positive constant.

Due to (2.10-2.13) the components of the electromagnetic energy-momentum four-vector and the three-vector of the spin angular momentum can be decomposed in the following way:

$$P^0 = \frac{1}{8\pi c} \int \left(\underline{E}^2 + \underline{B}^2\right) d^3x = \\ = \frac{1}{8\pi c} \int \left(\underline{E}_1^2 + \underline{B}_1^2\right) d^3x + \frac{1}{8\pi c} \int \left(\underline{E}_2^2 + \underline{B}_2^2\right) d^3x, \tag{2.14}$$

$$\underline{P} = \frac{1}{4\pi c} \int \underline{E} \times \underline{B} \cdot d^3x = \\ = \frac{1}{4\pi c} \int \underline{E}_1 \times \underline{B}_1 d^3x + \frac{1}{4\pi c} \int \underline{E}_2 \times \underline{B}_2 \cdot d^3x, \tag{2.15}$$

$$\underline{S} = \frac{1}{4\pi c} \int \underline{E} \times \underline{A} \, d^3x = \\ = \frac{1}{4\pi c} \int \underline{E}_1 \times \underline{A}_1 d^3x + \frac{1}{4\pi c} \int \underline{E}_2 \times \underline{A}_2 d^3x. \tag{2.16}$$

## 3. *A Scalar Constant of Motion*

Having in mind the decomposition (2.10-2.13) we can take the four-vectors $A_1^\alpha$ and $A_2^\alpha$ of two free electromagnetic fields and belonging electromagnetic field strengths $F_1^{\alpha\beta}$ and $F_2^{\alpha\beta}$ in order to construct the four-vector (3.1)

$$j^\alpha = F_1^{\alpha\beta} A_{2\beta} - F_2^{\alpha\beta} A_{1\beta}. \qquad (3.1)$$

Due to (2.5-2.8) this four-vector satisfies the continuity equation

$$\partial_\alpha j^\alpha = 0 \qquad (3.2)$$

from which the scalar constant of motion is obtained

$$Q = const._q \int \left( F_1^{0\mu} A_{2\mu} - F_2^{0\nu} A_{1\nu} \right) d^3x \qquad (3.3)$$

i.e.

$$Q = const._q \int \left( \underline{E}_1 \underline{A}_2 - \underline{E}_2 \underline{A}_1 \right) d^3x. \qquad (3.4)$$

It is worth to notify that the physical dimensions of the integrands in (2.16) and (3.4) are the same.

## 4. Electromagnetic Field Particles

Let the plane wave be moving in the positive z-direction. Then three-vector $\underline{A}$ is in the *xy*-plane and it satisfies the wave equation

$$\partial_\alpha \partial^\alpha \underline{A} = 0, \qquad (4.1)$$

with the condition (Coulomb gauge)

$$A^0 = 0, \quad div\underline{A} = 0. \qquad (4.2)$$

According to (2.10-2.13) it can be decomposed into two orthogonal three-vectors $\underline{A}_1$ and $\underline{A}_2$ those also lie in the *xy*-plane and have equal absolute values

$$|\underline{A_1}| = |\underline{A_2}| = a, \quad \frac{1}{2}\underline{A}^2 = a^2. \tag{4.3}$$

Two specific cases of monochromatic plane waves will be considered.

*Case (a):*

$$\underline{A} = a\left[\cos(\omega t - kz) - \sin(\omega t - kz)\right]\underline{i} + \\ + a\left[\sin(\omega t - kz) + \cos(\omega t - kz)\right]\underline{j}, \tag{4.4}$$

$$\underline{A_1} = a\left[\cos(\omega t - kz)\underline{i} + \sin(\omega t - kz)\underline{j}\right], \tag{4.5}$$

$$\underline{A_2} = a\left[-\sin(\omega t - kz)\underline{i} + \cos(\omega t - kz)\underline{j}\right], \tag{4.6}$$

$$\underline{A_1} \times \underline{A_2} = a^2 \underline{n}, \quad (\underline{n} = \underline{i} \times \underline{j}). \tag{4.7}$$

The energy-momentum four-vector and the three-vector of the spin angular momentum for this electromagnetic wave are given by $(k^\alpha = (k, \underline{k}))$

$$P^\alpha = \frac{1}{2\pi c} a^2 k L^3 k^\alpha, \tag{4.8}$$

$$\underline{S} = \frac{1}{2\pi c} a^2 k L^3 \underline{n}. \tag{4.9}$$

The scalar constant of motion reads

$$Q = const_q \left(-2a^2 k L^3\right). \tag{4.10}$$

Selecting $const_q = \frac{-1}{4\pi c}$ and denoting the scalar constant of motion by $\theta$ one gets

$$P^\alpha = \theta k^\alpha, \tag{4.11}$$

$$\underline{S} = \theta \underline{n}. \tag{4.12}$$

Since the quantity $\theta$ is a relativistic scalar this result is essentially equal to the Einstein's hypothesis. Therefore, the photon-like particles are possible physical objects already in the classical electromagnetic field theory.

The vectors $\underline{A}_1, \underline{A}_2, \underline{n}$ form right-hand oriented set of orthogonal vectors with right-circular polarization (as seen in the direction of propagation). The state of spin is +1.

*Case (b):*

$$\underline{A} = a\left[\cos(\omega t - kz) - \sin(\omega t - kz)\right]\underline{i} \\ + a\left[-\sin(\omega t - kz) - \cos(\omega t - kz)\right]\underline{j}, \quad (4.13)$$

$$\underline{A}_1 = a\left[\cos(\omega t - kz)\underline{i} - \sin(\omega t - kz)\underline{j}\right], \quad (4.14)$$

$$\underline{A}_2 = a\left[-\sin(\omega t - kz)\underline{i} - \cos(\omega t - kz)\underline{j}\right], \quad (4.15)$$

$$\underline{A}_1 \times \underline{A}_2 = a^2(-\underline{n}), \quad (\underline{n} = \underline{i} \times \underline{j}). \quad (4.16)$$

The energy-momentum four-vector and the three-vector of the spin angular momentum for this electromagnetic wave are given by

$$P^\alpha = \theta k^\alpha, \quad (4.17)$$

$$\underline{S} = \theta(-\underline{n}). \quad (4.18)$$

The scalar constant of motion is the same as in the case (a). This "photon" has the spin state -1.

The vectors $\underline{A}_2, \underline{A}_1, \underline{n}$ form right-hand oriented set of orthogonal vectors with left-circular polarization.

## 5. *Conclusion*

This simple analysis shows that photon-like particles are not strange within the conceptual framework of the classical electromagnetic field theory. Circular polarized waves lead to photons. Thus, light quantum hypothesis is not necessary.